\documentclass{nature}
\usepackage{graphicx}
\newcommand{\beginsupplement}{%
        \setcounter{table}{0}
        \renewcommand{\thetable}{S\arabic{table}}%
        \setcounter{figure}{0}
      	\renewenvironment{figure}{\let\caption\supcaption}{}	
	\newcommand{\supcaption}{
	    	\refstepcounter{figure}
    		\ifthenelse{\value{figure}=1}{
        		\noindent
    		}{
        		\par
   		 }
    		\sffamily\noindent\textbf{Figure S\arabic{figure}}\hspace{1em}
		}
		
	\renewcommand{\thefigure}{S\arabic{figure}}%
     	}

\title{Persistent spin excitations in doped antiferromagnets revealed by resonant inelastic light scattering}

\author{C.~J.~Jia$^{1, 2}$, E.~A.~Nowadnick$^{1, 3}$, K.~Wohlfeld$^{1}$, Y.~F.~Kung$^{1, 3}$, C.-C.~Chen$^{4}$, S. Johnston$^{5,6}$, T.~Tohyama$^{7}$, B.~Moritz$^{1,8}$, T.~P.~Devereaux$^{1}$}

\begin{document}

\maketitle

\begin{affiliations}
 \item Stanford Institute for Materials and Energy Sciences, SLAC National Accelerator Laboratory and Stanford University, Menlo Park, California 94025, USA 
 \item Department of Applied Physics, Stanford University, Stanford, California 94305, USA
 \item Department of Physics, Stanford University, Stanford, California 94305, USA
 \item Advanced Photon Source, Argonne National Laboratory, Lemont, Illinois 60439, USA
 \item Department of Physics and Astronomy, University of British Columbia, Vancouver, British Columbia, Canada V6T 1Z1
 \item Quantum Matter Institute, University of British Columbia, Vancouver, British Columbia, Canada V6T 1Z4
 \item Yukawa Institute for Theoretical Physics, Kyoto University, Kyoto, 606-8502, Japan
 \item Department of Physics and Astrophysics, University of North Dakota, Grand Forks, North Dakota 58202, USA
\end{affiliations}

\begin{abstract}
How coherent quasiparticles emerge by doping quantum antiferromagnets is a key question in correlated electron systems underlying an understanding of the phase diagram of copper oxides. Recent resonant inelastic x-ray scattering (RIXS) experiments in hole-doped cuprates have purported to measure high energy collective spin excitations that persist well into the overdoped regime and bear a striking resemblance to those found in the parent compound, challenging the perception that spin excitations should weaken with doping and have a diminishing effect on superconductivity. Here we show that RIXS at the Cu $L_3$-edge indeed provides access to the spin dynamical structure factor once one considers the full influence of light polarization. Further we demonstrate that high-energy spin excitations do not correlate with the doping dependence of T$_c$, while low-energy excitations depend sensitively on doping and show 
ferromagnetic correlations. This suggests that high-energy spin excitations are marginal to pairing in cuprate superconductors.
\end{abstract}

\section*{Introduction}

Initial Cu $L_3$-edge RIXS measurements on undoped and weakly underdoped cuprates~\cite{Lucio2010, Guarise2010} complemented earlier neutron and Raman scattering experiments, seeming to favor a spin-fluctuation scenario as a viable explanation of superconductivity~\cite{Tacon2011}.
However, more recent RIXS measurements on overdoped cuprates~\cite{Tacon2011,Tacon2013, Dean2013a, Dean2013b} have shown persistent high-energy spin excitations to very high doping levels where superconductivity disappears. In contrast neutron and Raman measurements display an absence of robust spin excitations in the overdoped regime~\cite{Wakimoto2007, Yuan2012}; this conflict undermines an understanding of unconventional superconductivity 
in the cuprates, 
making an investigation of how spin excitations manifest in the RIXS cross section a crucial component to its resolution.

In this letter, we reconcile these seemingly incompatible experimental results by computing Cu $L_3$-edge RIXS spectra using exact diagonalization (ED), capable of reproducing major experimental features.  We demonstrate with light polarization analysis that the RIXS cross-section in a crossed-polarization geometry can be interpreted simply in terms of the spin dynamical structure factor $S(\mathbf{q}, \omega)$ which enables a comparison between different scattering experiments. Utilizing determinant quantum Monte Carlo (DQMC), we study in detail the momentum and doping dependence of $S(\mathbf{q}, \omega)$, finding strong changes near both $(\pi,\pi)$ and $(0,0)$ with relatively insensitive antiferromagnetic zone boundary (AFZB) paramagnons upon hole doping. Moreover, with electron doping these same AFZB paramagnons harden significantly, providing a testable experimental prediction from this work.  Underlying this observed behavior is a framework of local spin exchange which remains robust even with 
significant doping away from the parent antiferromagnet.  In contrast, our calculations show a sensitive evolution of low-energy paramagnons near $(0,0)$ and $(\pi,\pi)$ which give evidence for the predominance of ferromagnetic correlations. These results highlight the importance of spectral weight and dispersion at low energies in establishing a relevant energy scale and strength of spin fluctuations for pairing rather than higher-energy AFZB paramagnons.

\section*{Results}

\subsection{The relationship between RIXS and $S(\mathbf{q}, \omega)$}

\begin{figure}
\includegraphics[width=0.9\columnwidth]{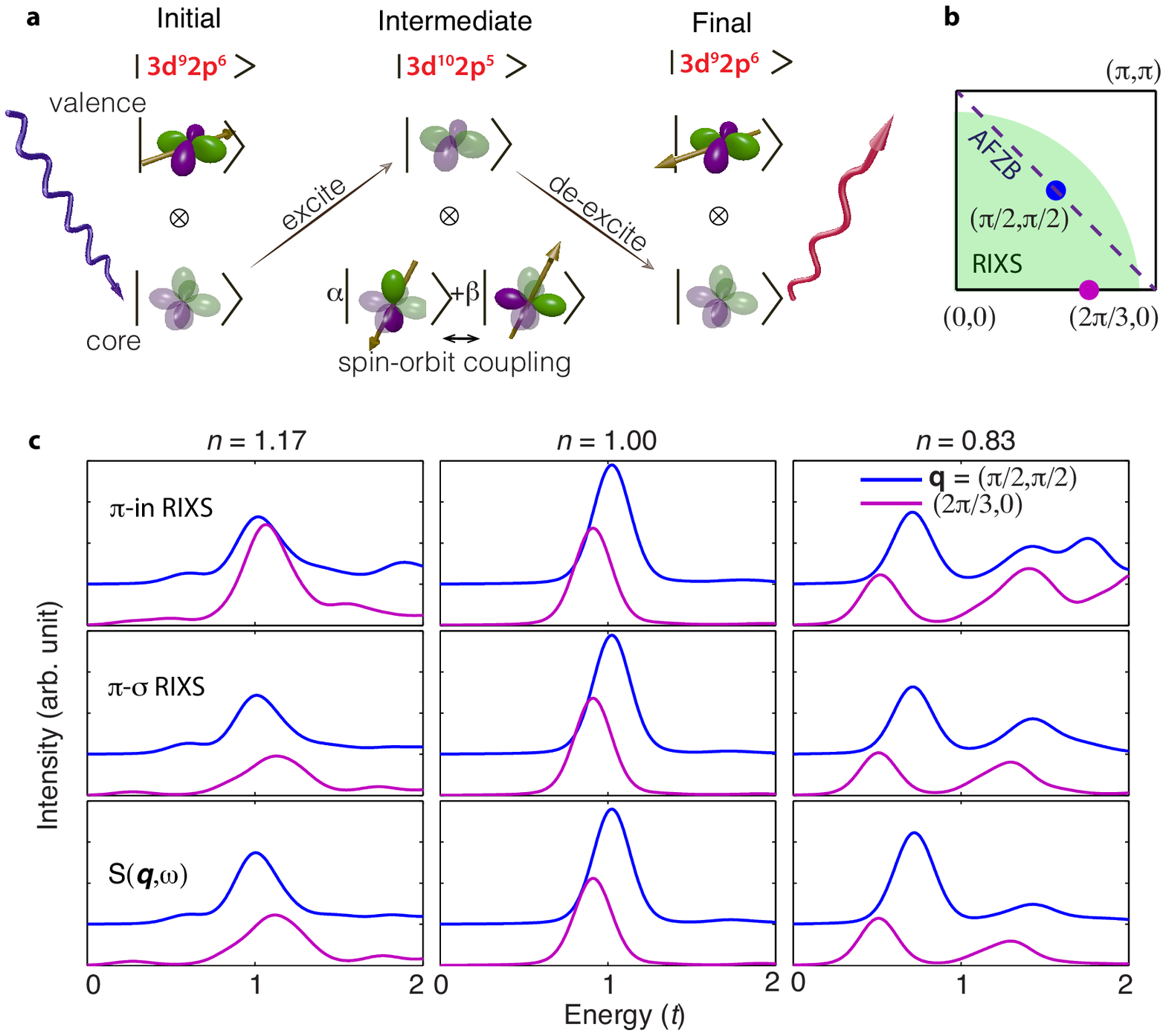}

\caption{\label{fig:RIXS}  
{\bf {Similarities between RIXS and the spin dynamical structure factor $S(\mathbf{q},\omega)$}}:
(a) Schematic diagram of a spin flip excitation produced during the RIXS process at the Cu $L_3$-edge.
The highlighted (darkened) orbitals represent holes. 
A spin-flip excitation is created when an electron with up spin is photoexcited from a Cu $2p$ core-level into the partially filled $3d_{x^2-y^2}$ orbital.  Subsequent de-excitation through the decay of an electron with spin down into the core produces a RIXS cross-section with a single spin-flip excitation.  Such a single spin-flip channel can only be enabled when the outgoing photon polarization is (or has non zero component) perpendicular to the incoming photon polarization. 
(b) A schematic picture of the area of the Brillouin zone accessible to RIXS at the Cu $L$-edge, as well as a line denoting the antiferromagnetic zone boundary (AFZB). The RIXS cross-section and the spin dynamical structure factor $S(\mathbf{q},\omega)$ have been evaluated in momentum space at the points marked by dots in the Brillouin zone [see also panel (c) below].
(c) The RIXS cross-section for select points in momentum space at the Cu $L_3$-edge (top and middle panels) compared against $S(\mathbf{q},\omega)$ (bottom panels).  Each has been calculated using exact diagonalization for the Hubbard model on a finite-size cluster for three different electron concentrations $n$. The top panels show RIXS spectra calculated for an in-coming polarization $\pi$ and a sum over the outgoing polarizations. The middle panels show the results with out-going polarization discrimination, here chosen in the cross-polarized geometry to emphasize the spin excitations. The results were obtained for the Hubbard model parameters $U=8t$, $t^{\prime}=-0.3t$ with $t=0.4eV$, a Lorentzian broadening with half width at half maximum (HWHM) = $0.025t$ and a Gaussian broadening with HWHM = $0.118t$ on the energy transfer. [See Methods] 
}
\end{figure}

RIXS is a resonant technique and its sensitivity to magnetic excitations arises as a result of core-level spin-orbit interactions in the intermediate state [see Fig.~\ref{fig:RIXS}(a)]. 
Although in Mott insulators it has been shown that the RIXS cross-section can be approximated by $S(\mathbf{q}, \omega)$ when the charge excitations are gapped, it is not clear whether the same approximation carries over to doped systems where the ground state is no longer that of a Mott insulator with commensurate filling~\cite{Ament2009, MauritsPRL}.

To answer this question, we numerically evaluate the RIXS cross-section as a function of momentum [Fig.~1(c)] directly from the Kramers-Heisenberg formula~\cite{RevModPhys.83.705,Jia2012} using small cluster ED of an effective single-band Hubbard Hamiltonian (including both nearest $t$ and next-nearest neighbor hopping $t'$ and on-site Coulomb repulsion $U$) at various electron concentrations $n$; details are given in the Methods section.
Figure \ref{fig:RIXS}($\textrm{c}$) displays the RIXS spectra calculated for the experimental geometry discussed in Ref.~\onlinecite{Tacon2011}. The RIXS spectra, even without outgoing polarization discrimination, agree well with $S(\mathbf{q}, \omega)$ for different electron concentrations at the chosen momentum space points accessible on the same finite size clusters for each calculation
. This is particularly true at half-filling where the charge gap ensures that only spin-excitations can be visible in the given energy range.  The main differences occur in the doped systems at higher energy (close to $2t$) which are of less interest for our spin analysis.  The important result shown in Fig.~1(c) concerns the suppression of these higher energy peaks in the cross-polarized geometry [see Fig.~1(c) $\pi-\sigma$ RIXS] which results in a significant improvement in the comparison between the RIXS cross-section and $S(\mathbf{q}, \omega)$.  This indicates that Cu $L_3$-edge RIXS with crossed polarizations (a four-particle correlator) provides access to the spin excitation spectrum encoded in $S(\mathbf{q}, \omega)$ (a two-particle correlator) for doped as well as undoped cuprates. [Ro further confirm this agreement between RIXS and $S(\mathbf{q}, \omega)$ for more momentum points, we can manually adjusted the Cu $L$-edge energy to be $1.8*930eV=1674eV$ so that momentum points up to $(\pi,\pi)$ can be reached. For more details see Supplementary Note 1]

\subsection{The momentum and doping dependence of $S(\mathbf{q}, \omega)$}

Having established a relationship between RIXS and $S(\mathbf{q}, \omega)$, we focus now on the momentum and doping dependence of $S(\mathbf{q}, \omega)$ for the single-band Hubbard model.  Here we employ the numerically exact DQMC method 
[see Methods]
with maximum entropy analytic continuation on larger lattices 
with fine control of the electron concentration through the chemical potential.
As shown in Fig.~\ref{fig:DQMC} 
the DQMC calculations qualitatively reproduce both the momentum and doping evolution of the RIXS measurements found in Refs.~\onlinecite{Lucio2010, Guarise2010, Tacon2011,Tacon2013, Dean2013a, Dean2013b}. A comparison between the intensity and dispersion of low-energy magnetic excitations near $(0,0)$ and $(\pi,\pi)$ shows a transition from antiferromagnetic to ferromagnetic spin correlations with increasing doping. In an antiferromagnetic system, the dynamical spin structure factors show gapless excitations at both $(0,0)$ and $(\pi,\pi)$, with strong intensity around $(\pi,\pi)$. In a ferromagnetic system, the dynamical spin structure factors show strong intensity with gapless excitations at $(0,0)$ and much weaker intensity with gapped excitations when approaching $(\pi,\pi)$. However, the spectra of higher-energy AFZB paramagnons show relatively little change with hole doping other than a general decrease of intensity, suggesting that spin excitations do not soften even in the heavily 
overdoped regime. For electron doping, AFZB paramagnons surprisingly harden by ~50\% at 15\% doping which has been observed recently in the prototypical electron-doped cuprate Nd$_{2-x}$Ce$_x$CuO$_4$.~\cite{WeiSheng} 
For additional analysis and discussion, see Supplementary Discussion.

\subsection{Theory to understand AFZB paramagnons}

The behavior of these AFZB paramagnons stands in stark contrast to naive expectations of spin softening with {\it either} hole or electron doping: (i) long-range AF order collapses quickly upon doping with a small (intermediate) concentration of holes (electrons), and one would expect spin excitations to soften accordingly~\cite{Vladimirov2009, Kar2011}; (ii) short-range AF correlations are further weakened due to a dilution of AF bonds [see Fig.~\ref{fig:cartoon}(a)] in the locally static spin picture. Indeed, the nearest neighbor spin-spin correlations from the DQMC calculations with electron doping decrease in a manner surprisingly well described by a locally static spin picture where the doped electrons are immobile, as shown in Fig.~3(b).  

We can address these points by considering the role of
three-site exchange~\cite{Daghofer2008} which lowers the system energy when both doped carriers and AF correlations are present [see Fig.~\ref{fig:cartoon}(a)].  A local spin flip in an otherwise AF background produces a ferromagnetic alignment of nearest neighbor spins which costs additional energy (of the same order as the spin exchange $J=4t^2/U$) by suppressing hole- (or double-occupancy-) delocalization represented by the three-site terms.  In fact, if we consider only the spin exchange contributions, the combined energy of a single spin flip in the doped system (breaking both spin exchange and three-site bonds) is larger than that of the undoped system by $\sim J/4$ [see  Fig.~\ref{fig:cartoon}(a)].  This hardening has been observed in ED calculations of $S(\mathbf{q}, \omega)$ for the $H_{Hubbard}$, $H_{t-J}$, and $H_{t-J}+H_{3s}$ Hamiltonians [see Methods] as shown in Fig.~\ref{fig:cartoon}($\textrm{c}$) upon electron doping.

\begin{figure}
\includegraphics[width=0.9\columnwidth]{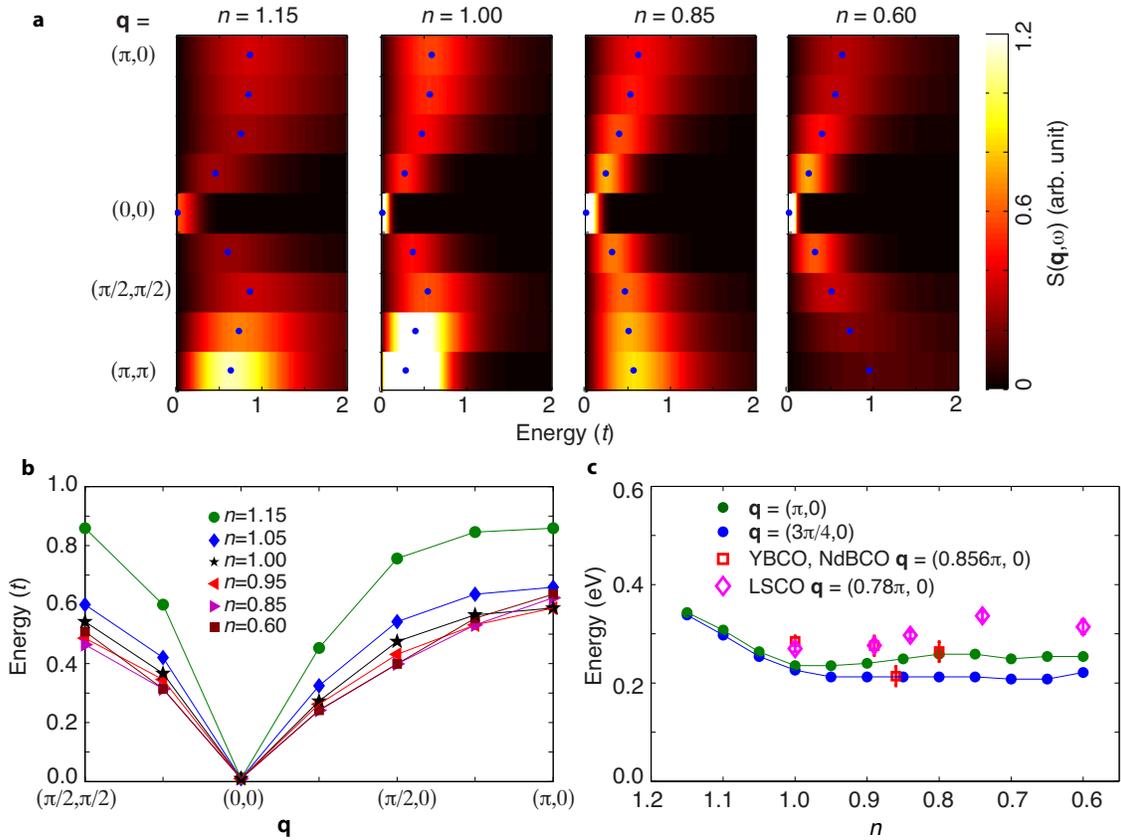}
\caption{\label{fig:DQMC} 
{\bf {The spin dynamical structure factor $S(\mathbf{q}, \omega)$ calculated using DQMC for the Hubbard model}}: (a) False color plots of the spectra along high symmetry directions in the Brillouin zone for different electron concentrations $n$. The calculations are done with the same Hubbard model parameters as in Fig.~1.
(b) Dispersion relations of the spin response peak in $S(\mathbf{q},\omega)$ along these high symmetry directions for different $n$.
(c) Comparison of the calculated and observed energies of the spectral peaks at selected momenta as a function of $n$ [$t=400$ meV has been used for the comparison]. YBa$_2$Cu$_3$O$_{6+x}$, YBa$_2$Cu$_4$O$_8$ and Nd$_{1.2}$Ba$_{1.8}$Cu$_3$O$_{6+x}$ RIXS experimental data are taken from~\onlinecite{Tacon2011}; La$_{2-x}$Sr$_x$CuO$_4$ RIXS experimental data are taken from~\onlinecite{Dean2013b}. }
\end{figure}

\begin{figure}
\includegraphics[width=0.9\columnwidth]{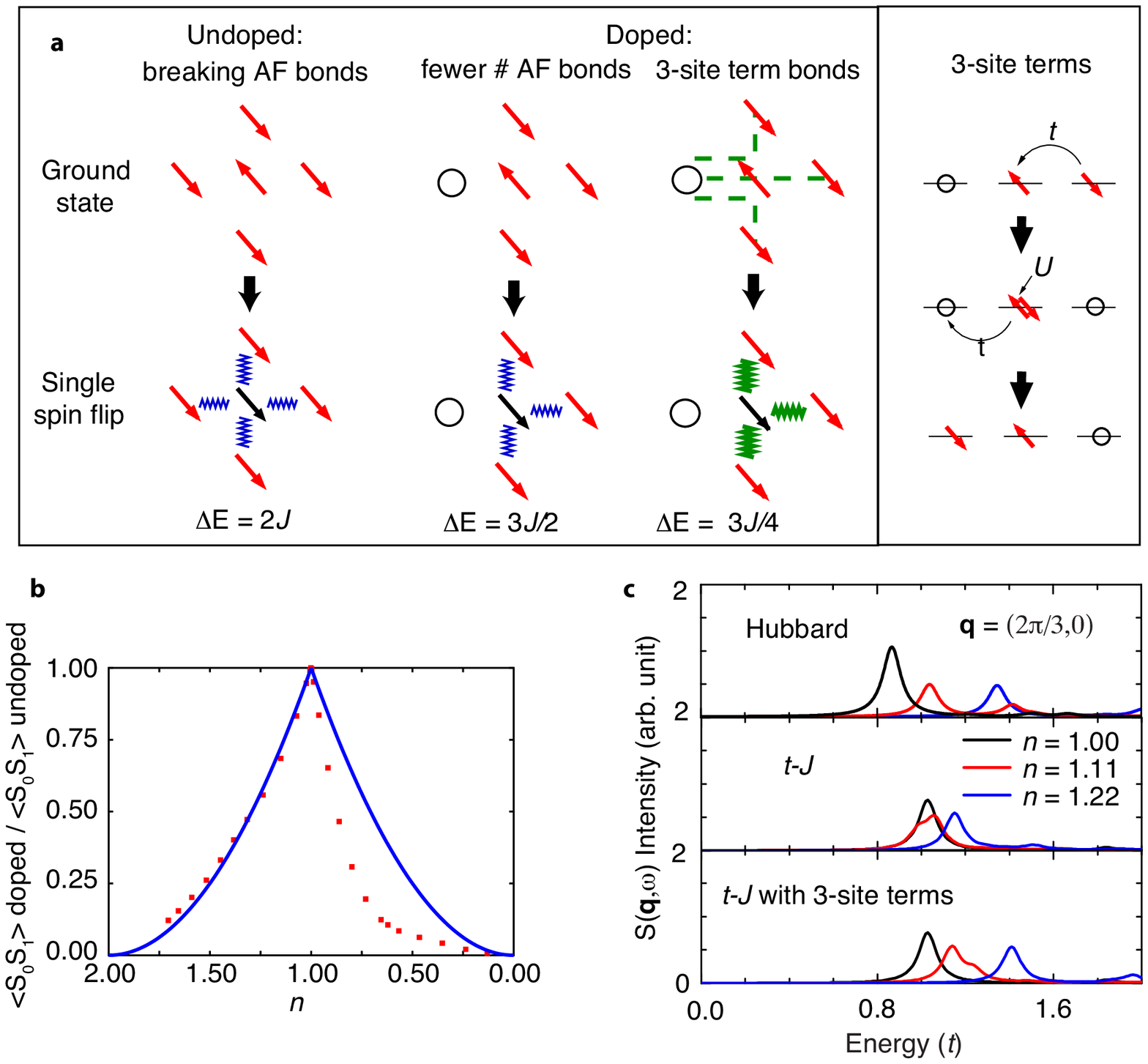}
\caption{ \label{fig:cartoon} 
{\bf {Understanding doping dependence of the spin dynamical structure factor $S(\mathbf{q},\omega)$}}:
(a) A cartoon illustrating the energy cost of spin excitations with doping within a ``locally static hole'' model: (top row) ground state with AF correlations; (bottom row) a single spin-flip excitation.  Green dashed lines represent the paths of hole delocalization by three-site terms, blue saw-tooth lines represent broken AF bonds, and green saw-tooth lines represent broken three-site bonds.  Undoped, a single spin-flip costs an energy of $2J$ from the four broken AF bonds.  With doping, this is reduced by the dilution of the AF background.  With the three-site terms the overall energy cost increases compared to the undoped system due to the reduction in hole-delocalization energy.  The side panel shows the process of hole-motion represented by the three-site terms, similar to the super exchange process. For electron doping, a particle-hole transformation can be applied so that a site with an open circle represents double occupancy. 
(b) Nearest neighbor spin-spin correlations $\langle {\bf S}_0 {\bf S}_1 \rangle$ as a function of electron concentration $n$ from DQMC.  The solid lines represent a fit of the doping dependence in the ``locally static hole" model. [see Methods]
(c) A comparison between ED results for $S(\mathbf{q}, \omega)$ in $H_{Hubbard}$, $H_{t-J}$, and $H_{t-J}+H_{3s}$ [see Methods] for different values of $n$ at $(2\pi/3, 0)$.  The three-site terms lead to hardening of spin excitations in qualitative agreement with the results from the Hubbard model.  We calculate $S(\mathbf{q}, \omega)$ on the three model Hamiltonians with the parameters $J = 0.4t$,
$t = 0.25t$, $U = 10t$ (corresponding to $J = 0.4t$ by the relation $J = 4t^2/U$) and a Lorentzian broadening HWHM = $0.05t$.
}
\end{figure}

The situation is more subtle with hole doping, because this ``locally static model" no longer completely applies as seen in Fig.~\ref{fig:cartoon}(b).
The negative next-nearest neighbor hopping $t'$ (positive for electron doping) promotes magnetic sublattice mixing and a much larger destruction of the AF correlations~\cite{Bala1995}.  With hole doping the trend observed in RIXS is fully recovered only in the Hubbard model, as shown in Fig.~\ref{fig:DQMC}, implying that higher order processes absent in $t$-$J$-type models become crucial in quantitatively reproducing the spin wave dispersion~\cite{Dellanoy2009}. 

\section*{Discussion}

\begin{figure}
   \includegraphics[width=0.9\columnwidth]{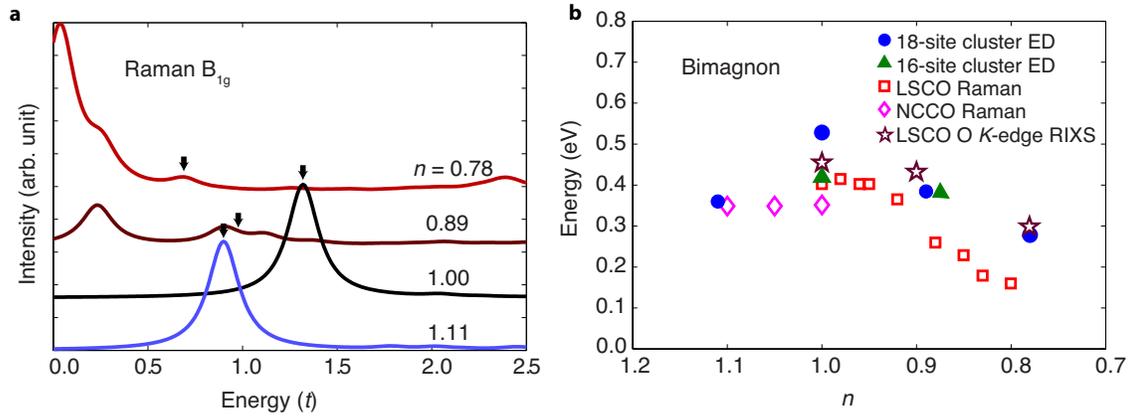}
  \caption{\label{fig:Raman} 
{\bf {Raman spectra as a function of $n$}}:
(a) The $B_{1g}$ Raman spectra for various values of $n$ calculated using exact diagonalization of the Hubbard model on 18-site clusters. [see Methods]. The calculations are done with the same Hubbard model parameters as in Fig.~1 and~2, and with a Lorentzian broadening HWHM = $0.1t$. Evolution of the spectral (`bimagnon') peak in the Raman spectra denoted by the black arrows [see Methods for identification of the `bimagnon' peaks] in  panel (a) together with the results from 16-site clusters (solid symbols). For the 18-site calculations for example, the bimagnon peaks are obtained by fitting the spectra around the black arrow in panel (a) with Lorentzian functions. These results are compared to experiments (open symbols) obtained from Raman scattering
on La$_{2-x}$Sr$_x$CuO$_4$~\onlinecite{Muschler2010} and Nd$_{2-x}$Ce$_x$CuO$_4$~\onlinecite{Onose2004}, and oxygen $K$-edge RIXS on La$_{2-x}$Sr$_x$CuO$_4$~\onlinecite{Bisogni2012}.} 
\end{figure}

How do these results reconcile the seemingly contradictory observations between RIXS, neutron and Raman scattering?  First, inelastic neutron scattering probes spin excitations particularly well around $(\pi, \pi)$ momentum transfer, showing a vanishing  spectral weight in the regime of large hole doping $p\simeq0.3$~\cite{Wakimoto2007, JPSJNeutronReview}. This behavior is also visible in the numerical results presented in Fig.~2(a) which suggest that the impact of doping on the intensity and dispersion of excitations near $(0,0)$ and $(\pi,\pi)$ is not symmetric.
The decreasing correlation length with doping, evidenced by the spin gap at $(\pi,\pi)$ and the weak dispersion towards $(\pi/2,\pi/2)$, thus impacts these momentum points more strongly than the AFZB paramagnons, in accordance with a locally static picture.
Second, Raman scattering~\cite{Yuan2012,Muschler2010,Onose2004} shows a softening of the so-called bimagnon (double spin-flip or two-magnon) response upon both hole and electron doping.  This trend has been reproduced by our ED calculations of the $B_{1g}$ Raman response shown in Fig.~\ref{fig:Raman} [see Methods for the calculation details and the verification of bimagnon peaks]. 
Strong magnon-magnon interactions reduce the bimagnon Raman peak energy from twice that of the single magnon bandwidth as determined by AFZB magnons and quickly reduce the overall intensity. 
Taken as a whole, 
our results provide a qualitative, and in some cases quantitative, agreement with the salient experimental features of neutron scattering, Raman, and RIXS measurements, suggesting that coherent propagating spin waves quickly disappear with the destruction of long-range AF order upon doping, while short-range, single spin-flip processes can survive to high doping levels as reflected in the evolution of $S(\mathbf{q}, \omega)$.

Full polarization control will allow RIXS to become an effective tool for directly observing spin dynamics along the AFZB, particularly noting the electron/hole doping differences.  Together with the dome shaped superconducting phase diagram, these results imply that AFZB spin fluctuations 
might
play a relatively minor role in the pairing mechanism, consistent with established experimental and numerical observations~\cite{Scalapino2012,Tallon, White_Scalapino_1999}. This calls into question a simple view of pairing which emphasizes only the spin exchange energy scale $J$.  
However, we suggest that a definitive resolution to this issue would come from
 future RIXS experiments along the BZ diagonal (out to  $(\pi/2, \pi/2)$) to illuminate the evolution from antiferro- to ferro-magnetic correlations, compare with neutron scattering results, and ultimately shed additional light on the intriguing mystery of cuprate high-temperature superconductivity.

\section*{Methods}

\subsection{Numerical techniques}

We use exact diagonalization (ED) to evaluate the RIXS cross-section from the Kramers-Heisenberg formula~\cite{RevModPhys.83.705}, spin dynamical structure factor $S(\mathbf{q},\omega)$, and Raman scattering cross-section~\cite{RevModPhys.79.175} on small clusters with periodic boundary conditions.  We employ a 12-site Betts cluster in evaluating the RIXS cross-section and $S(\mathbf{q},\omega)$ shown in Fig.~1.  The Raman scattering response shown in Fig.~4 has been evaluated on 16- and 18-site square (or diamond-shaped) clusters and the 18-site cluster was employed to evaluate $S(\mathbf{q},\omega)$ for $H_\mathrm{Hubbard}$, $H_{t-J}$ and $H_{t-J}+H_{3s}$ shown in Fig.~3.  The ED calculations for $H_\mathrm{Hubbard}$ are performed with the Parallel ARnoldi PACKage (PARPACK) and the cross-sections obtained by use of the biconjugate gradient stabilized method and continued fraction expansion~\cite{Jia2012}.  The ED calculations on $H_{t-J}$ and $H_{t-J}+H_{3s}$ models are performed using the Lanczos 
algorithm.  Finite temperature determinant quantum Monte Carlo (DQMC) simulations~\cite{Blank1981,White1989} were performed on $H_\mathrm{Hubbard}$ to obtain the imaginary time spin-spin correlation function from which the real frequency response function $S(\mathbf{q},\omega)$ was obtained by analytic continuation using the maximum entropy method (MEM).~\cite{Jarrell1996,MEM2}  These simulations were performed on $8 \times 8$ square lattice clusters with periodic boundary conditions at an inverse temperature $\beta=3/t$ for the same Hubbard Hamiltonian parameter values utilized in the ED studies.  For this set of parameters, the DQMC method exhibits a significant fermion sign problem~\cite{SignProblemPRB} over the entire doping range which we address in the MEM~\cite{MEM2} (see Supplementary Methods and Supplementary Fig. 6). MEM requires the use of a model function for determining an entropic prior in the analytic continuation routine.  We utilize a Lorentzian model whose peak as a function of $\mathbf{q}$ is determined from a simple spin wave dispersion at 
small $\mathbf{q}$ out to the AFZB; however, beyond the AFZB the model assumes no softening as expected for long-range antiferromagnetism with the top of the magnon band set by approximations for the spin exchange $J$ and an assumed reduction of the spin moment by quantum fluctuations.  While some quantitative changes occur with significant changes to these default models, we have checked that the qualitative behavior remains robust.  The MEM routine returns the real frequency spin susceptibility from which $S(\mathbf{q},\omega)$ is obtained from the fluctuation-dissipation theorem. More details about the models and numerical algorithms can be found in the following Methods and Supplementary Methods.

\subsection{RIXS}
 
The Cu $L_3$-edge RIXS cross-section is calculated using the Kramers-Heisenberg formula~\cite{RevModPhys.83.705} for the single-band Hubbard model 

\begin{equation} 
\begin{aligned}
&I(\mathbf{q}, \omega, \omega_{in})
= \frac{1}{\pi} \mathrm{Im} \langle \Psi | \frac{1}{\mathit{H}_\mathrm{Hubbard} - \mathit{E}_0 -\omega - \mathit{i} 0^+} | \Psi \rangle \\
&| \Psi \rangle = \sum_{i\alpha\sigma} e^{i\mathbf{q}\cdot\mathbf{R}_{i}} \mathit{D^{\dagger}_{i\alpha\sigma}} \frac{1}{\mathit{H}_\mathrm{Hubbard}+\mathit{H}_{CH}-\mathit{E}_0-\omega_{in}-\mathit{i}\Gamma} \mathit{D_{i\alpha\sigma}} | 0 \rangle 
\end{aligned}
\end{equation} 
in which
\begin{equation} 
\begin{aligned}
&\mathit{H_\mathrm{Hubbard}}= -t \sum_{<ij>,\sigma} d_{i\sigma}^{\dagger}d_{j\sigma}-t^{\prime} \sum_{\ll ij\gg,\sigma} d_{i\sigma}^{\dagger}d_{j\sigma}+\sum_{i} U n_{i\uparrow}^d n_{i\downarrow}^d \\
&\mathit{H}_{CH}=\sum_{i\alpha\sigma}(\epsilon^d-\epsilon^p)(1-n_{i\alpha\sigma}^p)
-\mathit{U}_c\sum_{i\alpha\sigma\sigma^{\prime}} n_{i\sigma}^d (1-n_{i\alpha\sigma^{\prime}}^p) +\lambda\sum_{i\alpha\alpha^{\prime} \sigma\sigma^{\prime}} p_{i\alpha\sigma} ^{\dagger} \chi_{\alpha\alpha^{\prime}}^{\sigma\sigma^{\prime}} p_{i\alpha^{\prime}\sigma^{\prime}},
\end{aligned}
\end{equation} 
where $\mathbf{q}$ is the momentum transfer; $\omega_{in}$ and $\omega = \omega_{in} - \omega_{out}$ are the incident photon energy (in our study the Cu $L_3$-edge) and photon energy transfer, respectively; $\mathit{E}_0$ is the ground state energy of the system in the absence of a core-hole; $| 0 \rangle$ is the ground state wave 
function; $\mathit{D_{i\alpha\sigma}} = \langle d_{x^2-y^2, \sigma} | \hat{\epsilon} \cdot \hat{r} | p_{\alpha\sigma} \rangle d^{\dagger}_{i\sigma}p_{i\alpha\sigma}$ (and h.c.) dictates the dipole transition process 
from Cu $2p$ to the $3d$ level (or from Cu $3d$ to $2p$), with the x-ray polarization $\hat{\epsilon}$ either $\pi$ or $\sigma$ (the polarization vector parallel or perpendicular to the scattering plane); and 
$\Gamma$ is the inverse core-hole lifetime (see Supplementary Note 2). In $\mathit{H_\mathrm{Hubbard}}$, $<...>$ and $\ll ... \gg$ represent a sum over the nearest and next nearest neighbor sites respectively. The Hamiltonian for the intermediate state also involves the on-site energy $\epsilon^d-\epsilon^p$ for creating a $2p$ 
core hole, Coulomb interaction $U_c$ induced by the core-hole and spin-orbit coupling $\lambda$, all denoted as in $\mathit{H}_{CH}$. $\chi_{\alpha\alpha^{\prime}}^{\sigma\sigma^{\prime}}\equiv\langle p_{\alpha\sigma}|\mathbf{l}\cdot \mathbf{s} |p_{\alpha^{\prime}\sigma^{\prime}}\rangle$ represents the spin-orbital coupling coefficients. The angle between the incident and the scattered photon propagation vectors is set to be $50^\circ$. The parameters used in the RIXS calculation are  $t=0.4$ eV, $U = 8t=3.2$ eV, $t^{\prime}=-0.3t=-0.12$ eV, $\epsilon^d-\epsilon^p=930$ eV, $U_c = -4t = -1.6$ eV, $\lambda = 13$ eV and $\Gamma=1t = 0.4$ eV~\cite{Tsutsui2000,Kourtis2012}. RIXS spectra at half-filling are taken only at the Cu $L_3$ resonance, and upon doping at the resonance closest to the half-filling Cu $L_3$-edge resonant energy. The RIXS results were obtained for a Lorentzian broadening with half width at half maximum (HWHM) = $0.01eV$ ($0.025t$) and a Gaussian broadening with HWHM = $0.047eV$ ($0.118t$) on the 
energy transfer. The 
spin dynamical structure factor $S(\mathbf{q},\omega)$, discussed in the next section, for $H_\mathrm{Hubbard}$ was calculated using the same parameters to make comparison to our RIXS results.

\subsection{Spin Dynamical Structure Factor}

The spin dynamical structure factor is defined as \begin{equation}S(\mathbf{q},\omega)= \frac{1}{\pi} \mathrm{Im} \langle 0 | s_{-\mathbf{q}} \frac{1}{H - \mathit{E}_0 - \omega - \mathit{i}0^+} s_{\mathbf{q}} | 0 \rangle,\end{equation} where we have studied the $H_\mathrm{Hubbard}$, $H_{t-J}$ and $H_{t-J}+H_{3s}$ Hamiltonians:

\begin{equation}H_{t-J} = -t \sum_{<ij>,\sigma} \tilde{c}_{i\sigma}^{\dagger}\tilde{c}_{j\sigma}-t^{\prime} \sum_{\ll ij\gg,\sigma} \tilde{c}_{i\sigma}^{\dagger}\tilde{c}_{j\sigma}+J \sum_{<ij>} S_{i}\cdot S_{j},\end{equation} \begin{equation}H_{3s} = -\frac{J}{4}\sum_{<ij><ij^{\prime}>,j\ne j^\prime,\sigma} (\tilde{c}_{j^{\prime}\sigma}^{\dagger}\tilde{n}_{i-\sigma}\tilde{c}_{j\sigma}-\tilde{c}_{j^{\prime}\sigma}^{\dagger}\tilde{c}_{i-\sigma}^{\dagger}\tilde{c}_{i\sigma}\tilde{c}_{j-\sigma});\end{equation}
$E_0$ is the corresponding ground state energy of the model Hamiltonian; $s_{\mathbf{q}}=\sum_{\mathbf{k}} c_{\mathbf{k+q},\uparrow}^{\dagger} c_{\mathbf{k},\uparrow} - c_{\mathbf{k+q},\downarrow}^{\dagger} c_{\mathbf{k},\downarrow}$, $c_{\mathbf{k},\sigma}^{\dagger} = \frac{1}{\sqrt{N}}\sum_i c_{i\sigma}^{\dagger} e^{i\mathbf{k} \cdot \mathbf{R}_i}$ for $H_\mathrm{Hubbard}$; $s_{\mathbf{q}}=\sum_{\mathbf{k}} \tilde{c}_{\mathbf{k+q},\uparrow}^{\dagger} \tilde{c}_{\mathbf{k},\uparrow} - \tilde{c}_{\mathbf{k+q},\downarrow}^{\dagger} \tilde{c}_{\mathbf{k},\downarrow}$, $\tilde{c}_{\mathbf{k},\sigma}^{\dagger} = \frac{1}{\sqrt{N}}\sum_i \tilde{c}_{i\sigma}^{\dagger} e^{i\mathbf{k} \cdot \mathbf{R}_i}$ for $H_{t-J}$ and $H_{3s}$; $S_i = \frac{1}{2}\sum_{\sigma\sigma^{\prime}}\tilde{c}_{i\sigma}^{\dagger}\mathbf{\sigma}_{\sigma\sigma^{\prime}}\tilde{c}_{i\sigma^{\prime}}$; and $\tilde{c}_{i\sigma}$ is restricted in the subspace without double occupancy $\tilde{c}_{i\sigma}=\bar{c}_{i\sigma}(1-\bar{n}_{i-\sigma})$ 
and $c_{i\sigma}=U^\dagger\bar{c}_{i\sigma}U$, in which the operator $\bar{c}_{i\sigma}$ annihilates a dressed electron whose hopping conserves the 
number of effective doubly occupied sites~\cite{Harris1967}. To explore the similarities and differences between $H_\mathrm{Hubbard}$ and $H_{t-J}$ (with and without $H_{3s}$), we calculate $S(\mathbf{q},\omega)$ on the three model Hamiltonians with the parameters $J=0.4t$, $t^{\prime}=-0.25t$ and $U=10t$ (corresponding to $J=0.4t$ by the relation $J=4t^{2}/U$). 
 
 \subsection{Locally Static Model}
 
In the ``locally static model" [see Fig.~3(b)] it is assumed that the holes destroy the short-range
spin-spin correlations solely by the effect of `static' doping, i.e. by removing
the spins and thus cutting spin bonds. In this case the nearest neighbor spin-spin
correlation can be calculated in the following way:
\begin{align}
\langle {\bf S}_0 {\bf S}_1 \rangle_{doped} \simeq (1-p)^2  \langle {\bf S}_0 {\bf S}_1 \rangle_{undoped},
\end{align} 
where $\langle {\bf S}_0 {\bf S}_1 \rangle$ is the abbreviation of $\langle 0 | {\bf S}_{i} {\bf S}_{j} | 0 \rangle$ for two neighboring sites $i$ and $j$, and $p$ is the concentration of either doped holes ($p=1-n$) or doped electrons ($p=n-1$).

 \subsection{Raman Scattering}
 
We calculate the Raman scattering cross-section in the $B_{1g}$ channel using the non-resonant response function for $H_\mathrm{Hubbard}$~\cite{RevModPhys.79.175}: 
\begin{equation}
\begin{aligned}
&\mathit{R}_{B1g}(\omega)= \frac{1}{\pi} \mathrm{Im} \langle 0 | \gamma_{B_{1g}} \frac{1}{H - \mathit{E}_0 - \omega - \mathit{i}0^+} \gamma_{B_{1g}} | 0 \rangle,\\
&\gamma_{B_{1g}}=\frac{1}{t} \sum_\mathbf{k} \Big(\frac{\partial ^2 \varepsilon (\mathbf{k})}{\partial k_x^2 }-\frac{\partial ^2 \epsilon (\mathbf{k})}{\partial k_y^2}\Big)  c_{\mathbf{k}}^{\dagger}c_{\mathbf{k}} = \frac{1}{2}\sum_\mathbf{k}(cos(k_x)-cos(k_y))c_{\mathbf{k}}^{\dagger}c_{\mathbf{k}}
\end{aligned}\end{equation} 
in which $\varepsilon (\mathbf{k}) = -2t(cosk_x + cosk_y) - 4t^\prime cosk_x cosk_y$ is the bare band dispersion, with parameters $U=8t$ and $t^{\prime}=-0.3t$. $t$ is taken as $0.4$ eV to make comparison with experimental data. 

This two-particle response also has been studied recently in cluster dynamical mean-field theory~\cite{Millis2012} showing that, if calculated fully gauge invariantly, 
the non-resonant Raman $B_{1g}$ response shows the presence of a strong bimagnon peak at half filling.
Nevertheless, the Raman spectrum calculated using this method for doped systems is sensitive to both charge and spin excitations in the low energy regime. 
Our identification of the bimagnon excitations in the Raman spectra relies primarily on the qualitative evolution of the peaks in agreement with experimental observations~\cite{Muschler2010}.  We note that all of the excitations visible in our Raman spectra correspond to $\Delta S = 0$ transitions which have $B_{1g}$ symmetry.  At half-filling, the energy of the excitation to which we assign bimagnon character lies within the charge gap which makes the bimagnon assignment clear.  Upon either hole or electron doping, we expect to develop charge excitations in the Raman response at low energy and for the two-magnon response to soften and decrease in intensity.  Our assignment corresponds to an upper bound for the bimagnon energy scale with doping where the additional structure at low energies signals either charge excitations or a substantial broadening of the bimagnon excitations now represented by multiple features in the exact diagonalization (ED) result (as discussed in connection with comparisons between determinant quantum monte carlo (DQMC) and ED results). However, the energy scale clearly softens and, more importantly, the intensity drops (significantly) in agreement with the experimental observations where the bimagnon ``peak'' becomes nearly impossible to distinguish from the charge background almost immediately upon crossing the AFM phase boundary.






\begin{addendum}
 \item We would like to thank G. Ghiringhelli, R. Hackl, J. P. Hill, B. J. Kim, M. Le Tacon, W.-S. Lee and J. Tranquada for discussions. This work was supported at SLAC and Stanford University by the U.S. Department of Energy, Office of Basic Energy Sciences, Division of Materials Science and Engineering, under Contract No.~DE-AC02-76SF00515 and by the Computational Materials and Chemical Sciences Network (CMCSN) under Contract No.~DE-SC0007091. CJJ is also supported by the Stanford Graduate Fellows in Science and Engineering. CCC is supported by the Aneesur Rahman Postdoctoral Fellowship at Argonne National Laboratory, operated under the U.S. Department of Energy Contract No. DE-AC02-06CH11357. TT is supported by the Grant-in-Aid for Scientific Research (Grant No. 22340097) from MEXT. TT and TPD acknowledge the YIPQS program of YITP, Kyoto University. A portion of the computational work was performed using the resources of the National Energy Research Scientific Computing Center (NERSC) supported by the U.S. Department of Energy, Office of Science, under Contract No.~DE-AC02-05CH11231.
 \item[Author Information] The authors declare no competing financial interests. Correspondence and requests should be addressed TPD (tpd@stanford.edu).
\end{addendum}

\pagebreak

\pagebreak

\pagebreak

\beginsupplement

\section*{Supplementary Material: Persistent spin excitations in doped antiferromagnets revealed by resonant inelastic light scattering}






\section*{Supplementary Figures}

\begin{figure}
\includegraphics[width=0.65\columnwidth]{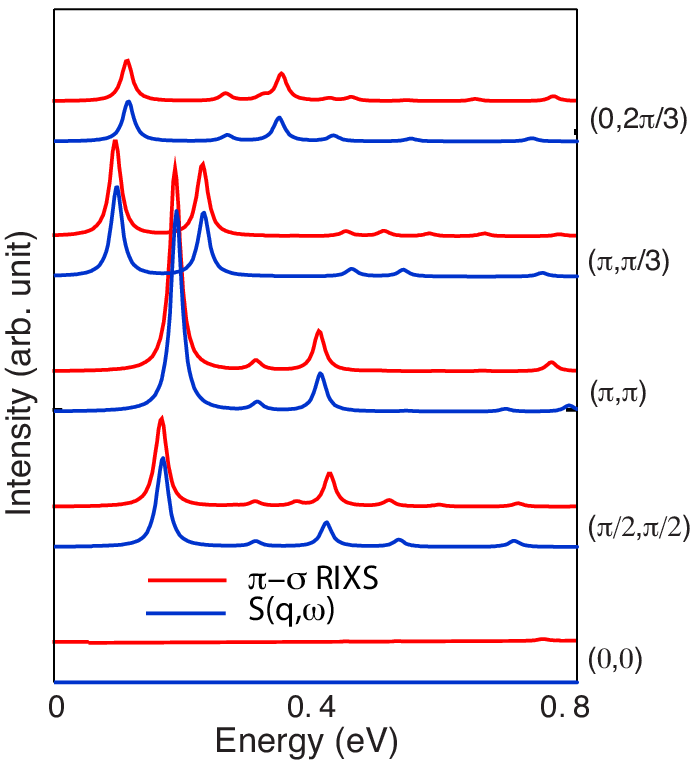}

\caption{\label{fig:supRIXS}  
{\bf The RIXS cross-section in the $\pi$-incoming/$\sigma$-outgoing polarization (red) and the spin dynamical structure factor $S(\mathbf{q},\omega)$ (blue):} The calculations are performed using exact diagonalization for the Hubbard model on a 12-site cluster for electron concentration levels $n = 0.83$. The Cu $L$-edge energy has been manually adjusted to be $1.8*930eV=1674eV$ so that momentum points up to $(\pi,\pi)$ can be reached. The other parameter values in this calculations are $U = 4eV,\ t = 0.37eV,\ t^{\prime} = -0.10 eV,\ U_c = 2eV,\ \Gamma = 0.4eV$. A Lorentzian broadening with HWHM = $0.01eV$ has been implemented for both RIXS and $S(\mathbf{q},\omega)$ calculations. The calculated spectra are shown with the energy in units of eV.}
\end{figure}

\pagebreak

\begin{figure}
\includegraphics[width=0.75\textwidth]{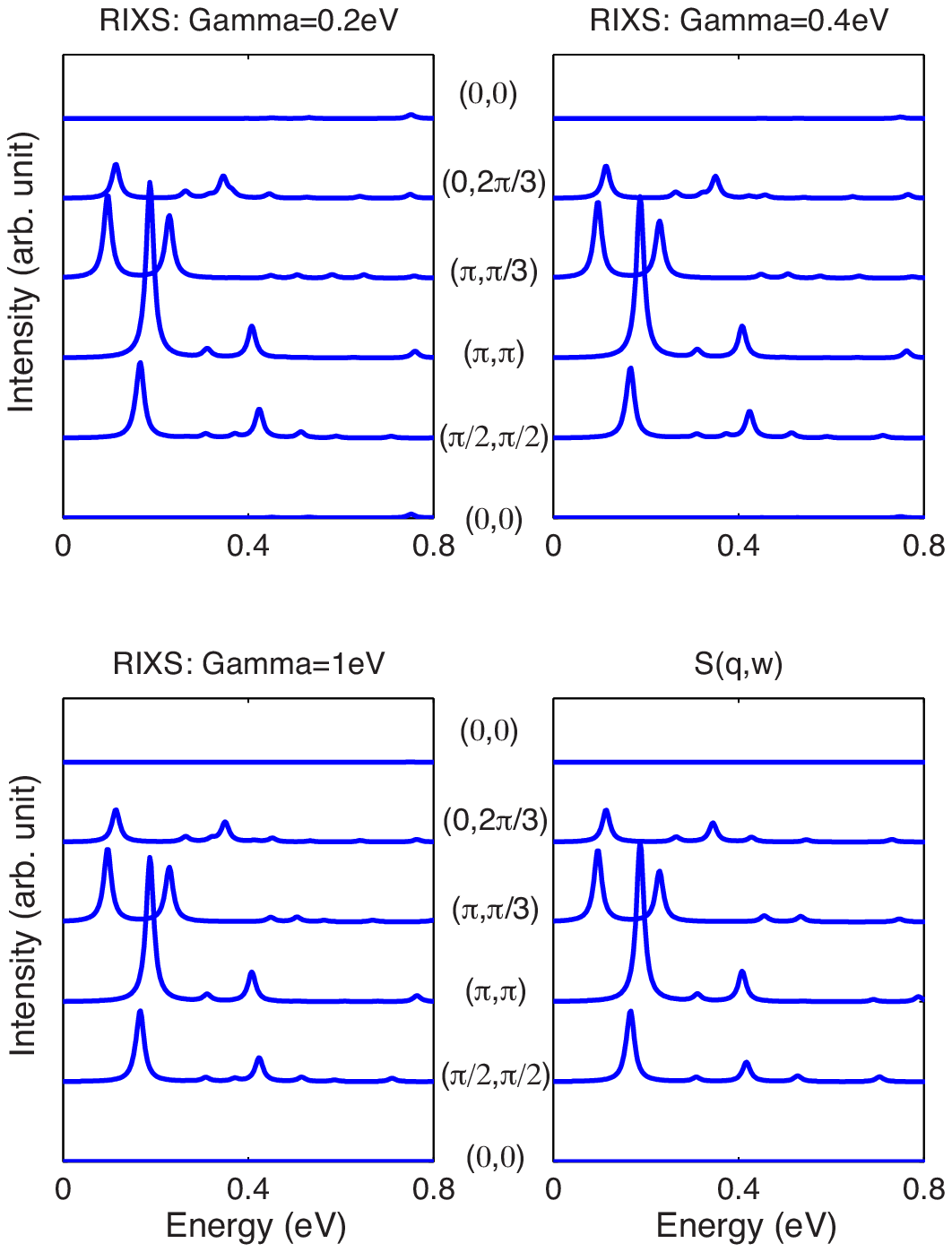}

\caption{ \label{fig:gamma} {\bf Core-hole lifetime $1/\Gamma$ dependence of Cu $L$-edge RIXS calculations vs. the dynamical structure factor $S(\mathbf{q},\omega)$}: The RIXS cross-sections at different momenta are shown for $\pi$-incoming/$\sigma$-outgoing polarization. The calculation parameters (other than $\Gamma$) and the electron concentration are the same as in Supplementary Fig.~\ref{fig:supRIXS}. }
\end{figure}

\begin{figure}
   \includegraphics[width=\columnwidth]{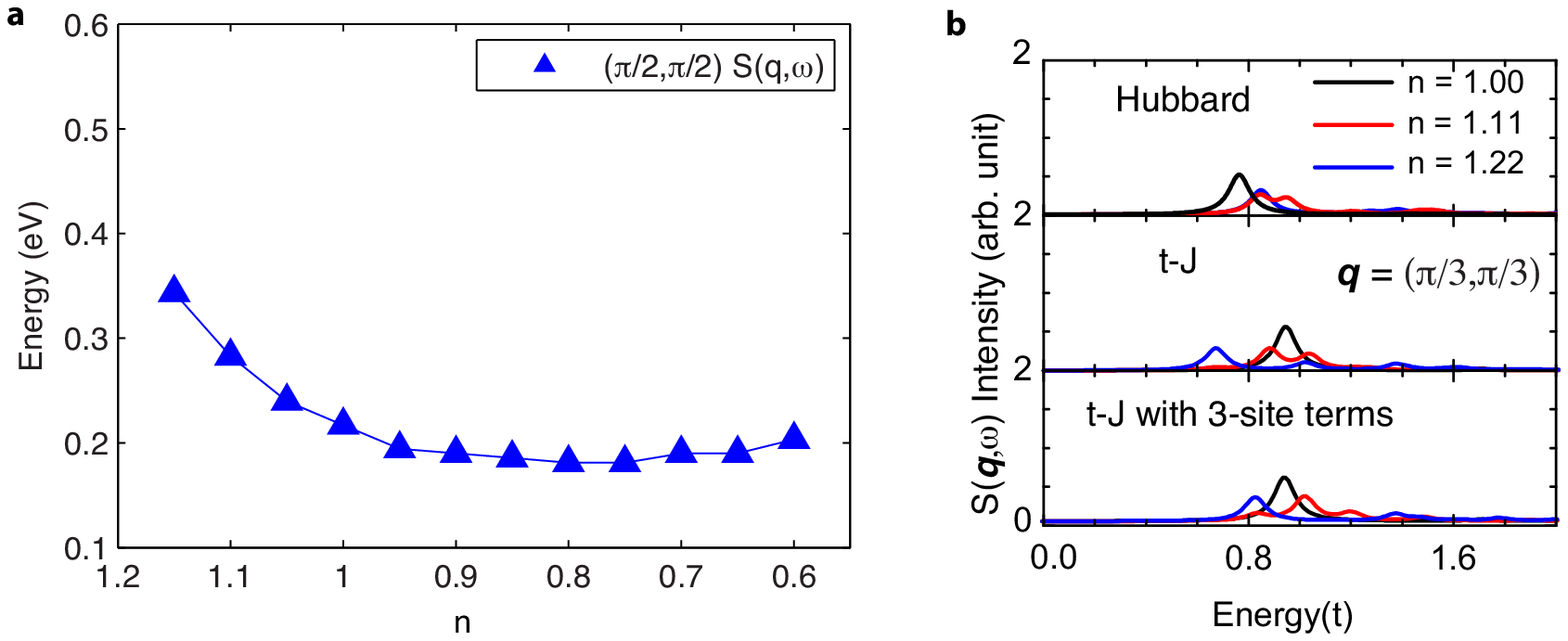}
  \caption{\label{fig:energy_pi2pi2} 
{\bf Peak energies of $S(\mathbf{q},\omega)$ at (around) $(\pi/2,\pi/2)$ as a function of doping}:
(a) Peak energies of $S(\mathbf{q},\omega)$ at $(\pi/2,\pi/2)$ as a function of the electron concentration obtained using DQMC in the single-band Hubbard model. The peak positions harden strongly upon electron doping, but show little change upon hole doping. (b) Exact diagonalization calculations of $S(\mathbf{q},\omega)$ on Hubbard, $t-J$ and $t-J$ plus 3-site terms models, for momentum $(\pi/3,\pi/3)$ (the momentum point closest to $(\pi/2,\pi/2)$ and accessible 
on the 18-site cluster). The calculation parameters are the same as in Fig. 3 of the main text. } 
\end{figure}

\begin{figure}
\includegraphics[width=\columnwidth]{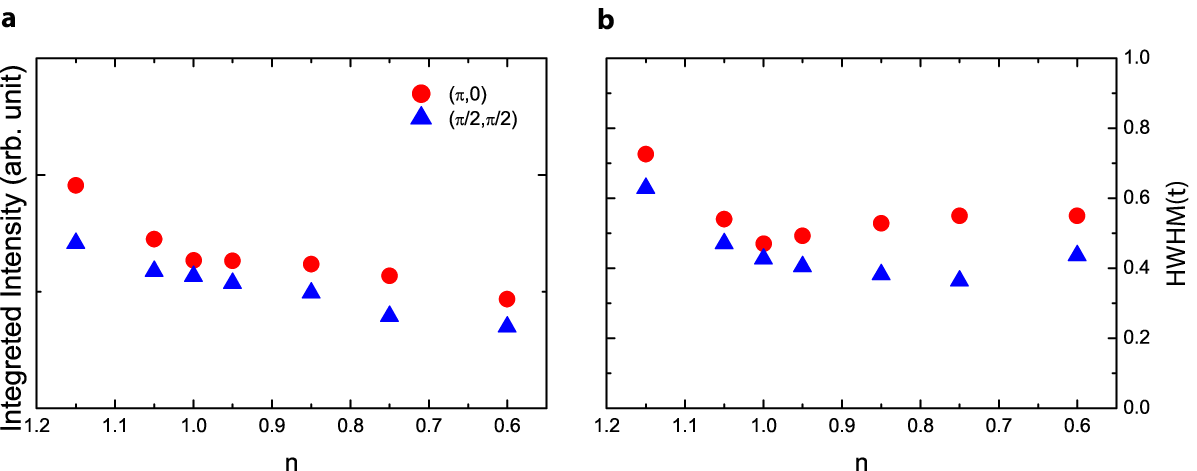}
\caption{ \label{fig:intensity_pi2pi2} 
{\bf Integrated intensity and HWHM for $S(\mathbf{q},\omega)$, $\mathbf{q}=(\pi,0)$ and $(\pi/2,\pi/2)$, as functions of electron concentration $n$. }
(a) Energy integrated intensity of $S(\mathbf{q},\omega)$ obtained by DQMC calculations for the single-band Hubbard model, with the same calculation parameters as in Fig.~2 of the main text. (b) Half-width at half-maximum (HWHM) obtained by fitting the same spectra with a single Gaussian waveform.} 
\end{figure}

\begin{figure}
\includegraphics[width=\columnwidth]{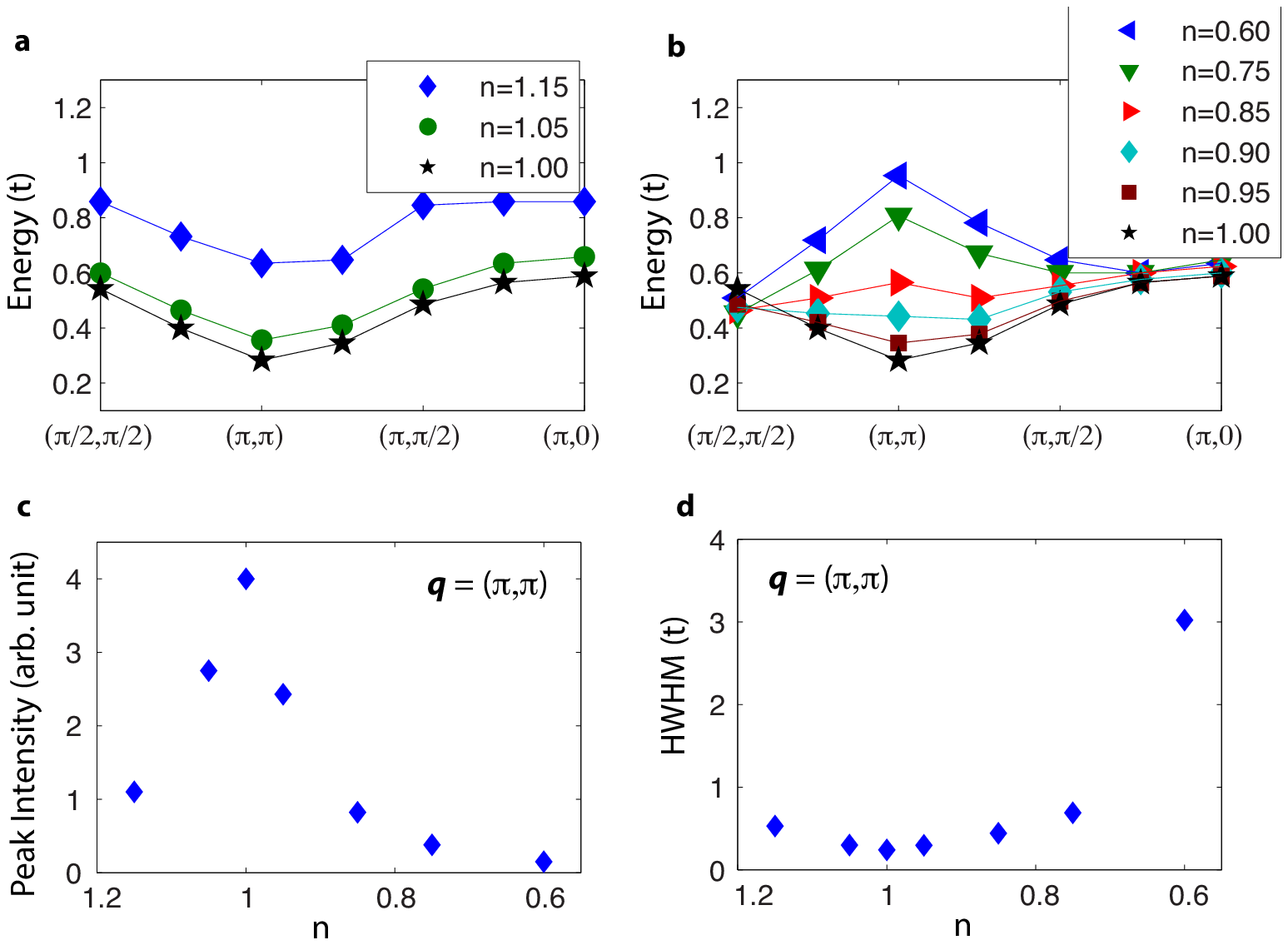}
\caption{ \label{fig:intensity} 
{\bf Dispersion relations of the spin response peak in $S(\mathbf{q},\omega)$ around $(\pi,\pi$) and its doping dependence}:
(a) Peak positions are extracted from the determinant quantum Monte Carlo calculations of $S(\mathbf{q},\omega)$ as shown in Fig.~2 of the main text. The dispersion is shown along high symmetry cuts, for multiple electron doping values. (b) Similar plots for hole doping. (c) The peak intensity of 
$S(\mathbf{q},\omega)$ at $\mathbf{q}=(\pi,\pi)$ obtained from DQMC. (d) HWHM for a single Gaussian peak fit to $S(\mathbf{q},\omega)$. At $n = 0.6$ or 40\% hole-doping, the spectrum becomes very broad and a second, broad peak emerges near the first, yielding a very large HWHM.}
\end{figure}

\begin{figure}
   \includegraphics[width=0.8\columnwidth]{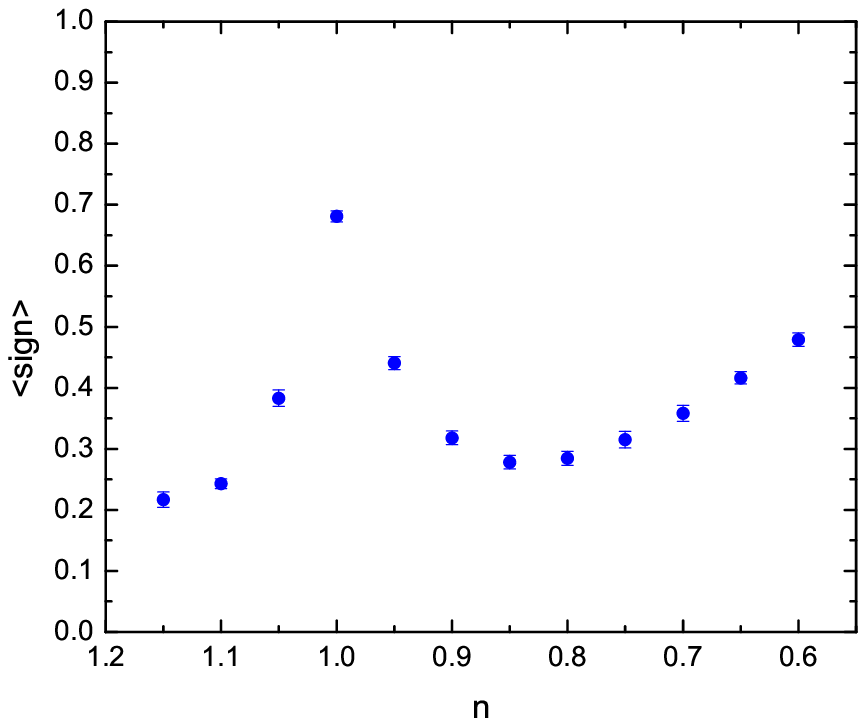}
  \caption{\label{fig:fermion} 
{\bf The fermion sign from the DQMC simulations as a function of doping}: the average fermion sign as well as the error bar of the fermion sign obtained form determinant quantum Monte Carlo calculations of $S(\mathbf{q},\omega)$ at multiple doping levels. The calculation parameters are the same as those used in Fig.~2 of the main text.
 } 
\end{figure}

\newpage

\begin{figure}
\includegraphics[width=0.7\columnwidth]{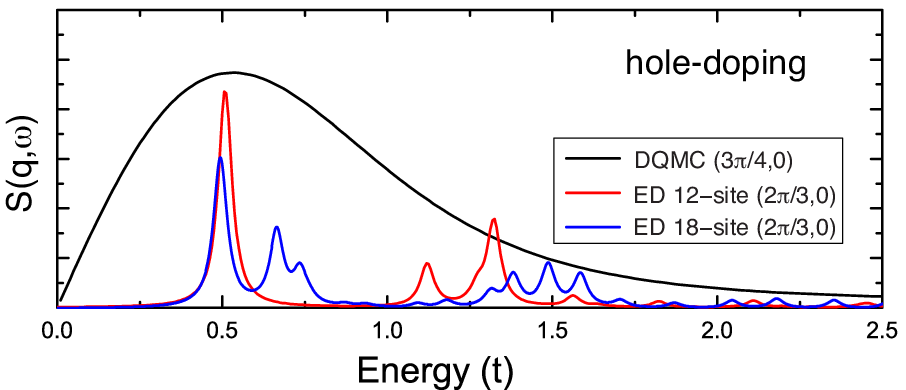}
\caption{\label{fig:supDQMC} 
{\bf Comparison of determinant quantum Monte Carlo and exact diagonalization calculations for $S(\mathbf{q}, \omega)$}: For DQMC, the electron density level is $n = 0.85$. For ED, the electron density level is two electrons less than half-filling: $n = 0.83$ for 12-site, and $n = 0.89$ for 18-site cluster. The calculations parameters are the same as in Fig.~1 of the main text. }
\end{figure}

\newpage

\section*{Supplementary Notes}

\subsection{Supplementary Note 1: The comparison of $S(\mathbf{q},\omega)$ with RIXS at other momentum points}

Showing more momentum points for comparison between RIXS and $S(\mathbf{q},\omega)$ would better demonstrate the agreement between the two spectra.  In the main text other momentum points provided by the largest available 12-site cluster has not been shown as these other points are not ``accessible'' in Cu $L$-edge RIXS due to the limited momentum transfer from photons at the Cu $L$-edge energy ($930eV$). Nevertheless, to answer solely questions about differences between $S(\mathbf{q},\omega)$ and RIXS, one can ``artificially" adjust the photon momentum by changing the Cu $L$-edge photon energy to cover the full Brillouin zone. Supplementary Fig.~\ref{fig:supRIXS} shows the RIXS cross-section in the $\pi$-incoming/$\sigma$-outgoing polarization (red) and the spin dynamical structure factor $S(\mathbf{q},\omega)$ (blue), with the Cu $L$-edge energy tuned to $1.8*930eV=1674eV$ for the remaining three momenta accessible on the 12-site cluster: $(\pi, \pi)$, $(\pi/3, \pi)$, and 
$(0, 0)$ (the latter is accessible in RIXS, but the intensity is nearly zero, so it was not shown in the main text).  Each spectrum has been calculated using exact diagonalization (ED) for the Hubbard model on a 12-site cluster for electron concentration $n = 0.83$. One can see that for each momentum, RIXS in the $\pi-\sigma$ geometry approximates well the spin dynamical structure factor $S(\mathbf{q},\omega)$. These results reconfirm the agreement between $S(\mathbf{q},\omega)$ and the RIXS cross-section in the cross-polarized scattering geometry.



\subsection{Supplementary Note 2: The core-hole lifetime dependence}

Our choice of $\Gamma = 0.4 eV$ in the main text is very realistic for cuprates -- in particular, it agrees with the most recent estimates in Ref.~\onlinecite{Bisogni2013} and is also in agreement with the chosen value of $\Gamma$ in Ref.~\onlinecite{Igarashi2012}. While the RIXS results might sensitively depend on $\Gamma$, choosing it in agreement with experiments properly captures all the effects related to finite core-hole lifetime. Nevertheless, decreasing or increasing the assumed value of $\Gamma$ by a factor of two {\it still} does not lead to significant differences in the spectral weight of excitations observed in the calculated RIXS spectrum.  More precisely, even for unphysical values of $\Gamma$, RIXS in the cross-polarized channel and in the doped case approximates fairly well the spin dynamical structure factor $S(\mathbf{q},\omega)$, as shown in Supplementary Fig.~\ref{fig:gamma}.
Naturally, further reduction of the value of $\Gamma$ (which definitely requires {\it unrealistic} values of $\Gamma$ at the Cu $L$-edge) leads to  significant changes in the RIXS spectra -- but the study of this phenomenon is beyond of the scope of this paper.

\section*{Supplementary Discussion}

\subsection{The behavior of $S(\mathbf{q},\omega)$ at $(\pi/2,\pi/2)$ and $(\pi,0)$ }

In Supplementary Fig.~\ref{fig:energy_pi2pi2} we show the behavior of the spin dynamical structure factor $S(\mathbf{q},\omega)$ at 
$(\pi/2,\pi/2)$ as a function of doping, in analogy with Fig.~2c in the main text. 
While the $(\pi/2,\pi/2)$ momentum point is hard to access in INS~\cite{JPSJNeutronReview} and has so far not been accessed in RIXS,
the spectrum at this point could be accessed and verified against these findings in future RIXS experiments. 

Let us note, that the behavior of $S(\mathbf{q},\omega)$ at $(\pi/2,\pi/2)$ as a function of doping is very similar to the behavior at $(\pi,0)$ 
as a function of doping (cf. left panel of Supplementary Fig.~\ref{fig:energy_pi2pi2} and Fig.~2c in the main text). However, unlike the $(\pi, 0)$ point, it is clear that the origin of the hole doping dependence can be explained less accurately using the $t-J$ model with 3-site terms (although the 3-site terms contribute to the hardening of the excitations with electron doping), cf. right panel of Supplementary Fig.~\ref{fig:energy_pi2pi2}. 
One must also invoke higher order exchange processes (present in the Hubbard model) to fully explain the observed doping dependence of these spin excitations.  We note that the ``local picture''  (see Fig. 3 of the main text) better represents the behavior of $S(\mathbf{q},\omega)$ 
along the $(0, 0)$-$(\pi, 0)$ direction rather than along the diagonal direction in the Brillouin zone due to a more significant destruction of the antiferromagnetic correlations close to $(\pi, \pi)$ with doping. 

For completeness, we show the doping dependence of the integrated intensity and the half width at half maximum (HWHM) at the $(\pi/2,\pi/2)$ and $(\pi, 0)$ points in Supplementary Fig.~\ref{fig:intensity_pi2pi2}.  Unlike the $(\pi, \pi)$ point (cf. Supplementary Fig.~\ref{fig:intensity} and discussion in the next section) the variation of intensity with doping at these momenta is rather small.  This agrees with the assumptions made when interpreting the RIXS results in terms of the ``locally static model'' and stating that one is primarily sensitive to local spin excitations at these momentum space points: the energy of these local excitations seems to be rather weakly affected by doping (see main text) and one also could expect that their intensity should not change very much with doping.

\subsection{The behavior of $S(\mathbf{q},\omega)$ near $(\pi,\pi)$ and the connection to INS}

Our calculated $S(\mathbf{q},\omega)$ is consistent with inelastic neutron scattering (INS), capturing important doping dependent characteristics of the measurements.  We find qualitative agreement between the DQMC results, as shown in Supplementary Fig.~\ref{fig:intensity}, and the INS measurements on the following points: 

On the hole doped side: (i) The momentum-integrated magnetic scattering intensity in a region around $(\pi,\pi)$ in the INS experiments strongly decreases with hole doping [cf. Fig. 7 (b) and Fig. 8 in Ref.~\onlinecite{JPSJNeutronReview}]; similarly, in our calculations the intensity of the magnetic response at $(\pi,\pi)$ and nearby points decreases significantly with hole doping. (ii) The energy scale of magnetic excitations near $(\pi,\pi)$ increases with significant hole doping~\cite{JPSJNeutronReview}, similar to the behavior we observe. 

On the electron doped side: (i) The momentum-integrated spectral weight of the so-called ``low energy peaks"  in the INS experiments seems to decrease more slowly with electron doping than with hole doping [cf. inset of Fig. 18 in Ref.~\onlinecite{JPSJNeutronReview}];  in our calculations we observe a similar trend -- the intensity of the $(\pi, \pi)$ peak decreases far slower with electron doping than with hole doping. (ii) The slope of low energy magnetic excitations near $(\pi, \pi)$ (which provides an estimate for the antiferromagnetic spin stiffness) becomes less steep with doping [cf. Fig. 18 in Ref.~\onlinecite{JPSJNeutronReview}]; we qualitatively observe a similar trend.
 
Nevertheless, the relatively high temperature and lack of a dense momentum-space mesh which are limiting factors of the DQMC method preclude any quantitative, substantive comparison between our results and some of the aspects of the INS spectra.  In particular, our momentum-space mesh does not provide sufficient resolution to address the magnetic incommensurability which is visible as part of the INS ``hourglass'' structure around $(\pi,\pi)$.

\section*{Supplementary Methods} 

\subsection{Fermion sign problem in DQMC}

The DQMC method for simulating the many-body Hubbard Hamiltonian exhibits a sign problem which restricts the minimum 
accessible temperatures (maximum $\beta$) and, to a lesser extent, the maximum accessible cluster sizes~\cite{SignProblemPRB}.  The sign problem arises from a non-positive definite measure of the probability distribution for different auxilliary field configurations in the DQMC method.  The sign problem can be partially overcome by performing measurements during DQMC process using a weighted distribution which accounts for this non-positive measure or ``fermion sign".  We have accounted for this source of statistical variation between different estimates of the correlation function using the maximum entropy method (MEM) analytic continuation techniques based on Bayesian inference~\cite{MEM2}, and the two-particle response can be reliably extracted for the values of the sign encountered in the simulations. (see Supplementary Fig.~6)

\subsection{ED vs DQMC}

Both ED and DQMC are numerically exact methods.  However, there are significant differences in the types of questions which can be better answered by each technique. 
Both methods are based on small, real-space cluster models of the full lattice Hamiltonian.  More specifically:

ED is a wavefunction-based method restricted by the exponential scaling of the Hilbert space dimension with cluster size which leads to two effects:  (i)  ED can be utilized only on small clusters.  The intermediate states of the Cu $L$-edge RIXS process for a 12-site cluster discussed in the main text, considering the Cu $2p$-core orbitals, has a Hilbert space dimension of $s$ = 14,976,864 at half-filling (while the size of the Hamiltonian matrix is $s \times s$). The Hamiltonian matrix is usually very sparse (usually a few hundreds of non-zero elements within each column or row), and can be handled by implementing massively parallel algorithms.   By extension, the size of the real-space cluster limits the number and spacing of accessible momentum space points for any comparison between RIXS and $S(\mathbf{q},\omega)$ in our study.  
(ii)  Any spectral function, e.g. single-particle addition or removal spectrum, spin and charge dynamical structure factors, RIXS cross-section, or Raman response function, only has finite spectral support over the discrete spectrum of eigenstates for the Hamiltonian represented in the given Hilbert space.  The spectral features thus naturally evolve, often significantly, with changes in the cluster size.  The advantages of ED over other techniques rest with the fact that it is a zero temperature formalism capable of addressing ground state properties and that it provides direct access to the wavefunction from which one can evaluate any correlation function or cross-section, i.e. RIXS and $S(\mathbf{q},\omega)$ on equal footing (albeit with severe limitations on the Hilbert space size). In particular, RIXS at the transition metal $K$-edge has been calculated by several groups using the ED technique~\cite{RevModPhys.83.705}.

DQMC is a finite temperature, imaginary time Green's function or propagator based method limited mainly by the fermion sign problem. One can perform simulations on significantly larger clusters with a more dense momentum space mesh than that of ED; however, the fermion sign problem limits the lowest temperatures that can be accessed with this method and one needs to employ an additional numerical procedure for obtaining real frequency response functions from the imaginary time data.  While we can obtain both single- and two-particle response functions using this method, there is to this point-in-time no method for estimating the multi-particle RIXS cross-section with this technique which requires an explicit handling of the core-hole intermediate state.  So, the DQMC method can provide full real-frequency information about single- and two-
particle response functions on a relatively dense mesh in momentum space, but limited to fairly high temperatures.

Despite their significant differences, these two methods give qualitatively similar results as shown in Supplementary Fig.~\ref{fig:supDQMC}. This figure displays a comparison between $S(\mathbf{q},\omega)$ obtained from DQMC simulations on $8\times8$ square clusters and that obtained from ED on both 12- and 18-site small clusters at approximately the same doping level and momentum space position.  One can see that the lowest energy magnon peak aligns in energy for both DQMC and ED. The presence of several discrete higher energy states in ED makes up part of the tail emanating on the high energy side of the main magnon peak in DQMC. In fact, increasing the ED cluster size shows that these states form a continuous band of excitations captured by DQMC.  This behavior is a manifestation of the discrete energy domain which supports spectral functions obtained using the ED technique.

The origin of the ``high energy" excitations, visible both in ED (as separate peaks at ca. $1.5t$ energy transfer) and DQMC (as the high energy ``tail'' extending beyond ca. $t$ energy transfer), may be explained as follows: (i) Due to the strong interaction between holes/electrons and spin fluctuations, there are many different ``spin-flip'' states in $S(\mathbf{q},\omega)$. (ii) These states can be understood as spin fluctuations dressed with holes/electrons (cf. Ref.~\onlinecite{Martinez1991} for a somewhat related problem of the hole dressed by magnons which also leads to spectral functions with ``many peaks"). It is remarkable that these dressed spin fluctuations have a very similar dispersion relation to that of the ``pure" spin fluctuations in the undoped case. The findings in this manuscript point to the origin of this effect in the peculiar interplay between the 3-site terms,  softening of spin excitations due to spin removal with hole/electron doping, and, most importantly, rather small effects on 
dispersion, at least on the electron doped side, due to the dressing of spin fluctuations with charge.

\section*{References}

\bibliography{biblio}

\begin{thebibliography}{10}
\expandafter\ifx\csname url\endcsname\relax
  \def\url#1{\texttt{#1}}\fi
\expandafter\ifx\csname urlprefix\endcsname\relax\def\urlprefix{URL }\fi
\providecommand{\bibinfo}[2]{#2}
\providecommand{\eprint}[2][]{\url{#2}}

\bibitem{Lucio2010}
\bibinfo{author}{Braicovich, L.} \emph{et~al.}
\newblock \bibinfo{title}{Magnetic excitations and phase separation in the
  underdoped {L}a$_{2-x}${S}r$_x${C}u{O}$_4$ superconductor measured by
  resonant inelastic x-ray scattering}.
\newblock \emph{\bibinfo{journal}{Phys. Rev. Lett.}}
  \textbf{\bibinfo{volume}{104}}, \bibinfo{pages}{077002}
  (\bibinfo{year}{2010}).

\bibitem{Guarise2010}
\bibinfo{author}{Guarise, M.} \emph{et~al.}
\newblock \bibinfo{title}{{Measurement of Magnetic Excitations in the
  Two-Dimensional Antiferromagnetic {S}r$_2${C}u{O}$_2${C}l$_2$ Insulator Using
  Resonant {X}-Ray Scattering: Evidence for Extended Interactions}}.
\newblock \emph{\bibinfo{journal}{Phys. Rev. Lett.}}
  \textbf{\bibinfo{volume}{105}}, \bibinfo{pages}{157006}
  (\bibinfo{year}{2010}).

\bibitem{Tacon2011}
\bibinfo{author}{{Le Tacon}, M.} \emph{et~al.}
\newblock \bibinfo{title}{{Intense paramagnon excitations in a large family of
  high-temperature superconductors}}.
\newblock \emph{\bibinfo{journal}{{Nature Physics}}}
  \textbf{\bibinfo{volume}{{7}}}, \bibinfo{pages}{725--730}
  (\bibinfo{year}{{2011}}).

\bibitem{Tacon2013}
\bibinfo{author}{Le~Tacon, M.} \emph{et~al.}
\newblock \bibinfo{title}{Dispersive spin excitations in highly overdoped
  cuprates revealed by resonant inelastic x-ray scattering}.
\newblock \emph{\bibinfo{journal}{Phys. Rev. B}} \textbf{\bibinfo{volume}{88}},
  \bibinfo{pages}{020501} (\bibinfo{year}{2013}).

\bibitem{Dean2013a}
\bibinfo{author}{Dean, M. P.~M.} \emph{et~al.}
\newblock \bibinfo{title}{{High-Energy Magnetic Excitations in the Cuprate
  Superconductor
  ${\mathrm{Bi}}_{2}{\mathrm{Sr}}_{2}{\mathrm{CaCu}}_{2}{\mathbf{O}}_{8\mathbf{+}\delta}$:
  Towards a Unified Description of Its Electronic and Magnetic Degrees of
  Freedom}}.
\newblock \emph{\bibinfo{journal}{Phys. Rev. Lett.}}
  \textbf{\bibinfo{volume}{110}}, \bibinfo{pages}{147001}
  (\bibinfo{year}{2013}).

\bibitem{Dean2013b}
\bibinfo{author}{Dean, M. P.~M.} \emph{et~al.}
\newblock \bibinfo{title}{{Persistence of magnetic excitations in
  La$_{2-x}$Sr$_x$CuO$4$ from the undoped insulator to the heavily overdoped
  non-superconducting metal}}.
\newblock \emph{\bibinfo{journal}{Nature Materials}}
  \textbf{\bibinfo{volume}{12}}, \bibinfo{pages}{1019--1023}
  (\bibinfo{year}{2013}).

\bibitem{Wakimoto2007}
\bibinfo{author}{Wakimoto, S.} \emph{et~al.}
\newblock \bibinfo{title}{{Disappearance of Antiferromagnetic Spin Excitations
  in Overdoped ${\mathrm{La}}_{2-x}{\mathrm{Sr}}_{x}{\mathrm{CuO}}_{4}$}}.
\newblock \emph{\bibinfo{journal}{Phys. Rev. Lett.}}
  \textbf{\bibinfo{volume}{98}}, \bibinfo{pages}{247003}
  (\bibinfo{year}{2007}).

\bibitem{Yuan2012}
\bibinfo{author}{Li, Y.} \emph{et~al.}
\newblock \bibinfo{title}{{Feedback Effect on High-Energy Magnetic Fluctuations
  in the Model High-Temperature Superconductor
  ${\mathrm{HgBa}}_{2}{\mathrm{CuO}}_{4+\delta}$ Observed by Electronic Raman
  Scattering}}.
\newblock \emph{\bibinfo{journal}{Phys. Rev. Lett.}}
  \textbf{\bibinfo{volume}{108}}, \bibinfo{pages}{227003}
  (\bibinfo{year}{2012}).

\bibitem{Ament2009}
\bibinfo{author}{Ament, L. J.~P.}, \bibinfo{author}{Ghiringhelli, G.},
  \bibinfo{author}{{Moretti Sala}, M.}, \bibinfo{author}{Braicovich, L.} \&
  \bibinfo{author}{van~den Brink, J.}
\newblock \bibinfo{title}{{Theoretical Demonstration of How the Dispersion of
  Magnetic Excitations in Cuprate Compounds can be Determined Using {R}esonant
  {I}nelastic {X}-{R}ay {S}cattering}}.
\newblock \emph{\bibinfo{journal}{Phys. Rev. Lett.}}
  \textbf{\bibinfo{volume}{103}}, \bibinfo{pages}{117003}
  (\bibinfo{year}{2009}).

\bibitem{MauritsPRL}
\bibinfo{author}{Haverkort, M.~W.}
\newblock \bibinfo{title}{Theory of resonant inelastic x-ray scattering by
  collective magnetic excitations}.
\newblock \emph{\bibinfo{journal}{Phys. Rev. Lett.}}
  \textbf{\bibinfo{volume}{105}}, \bibinfo{pages}{167404}
  (\bibinfo{year}{2010}).

\bibitem{RevModPhys.83.705}
\bibinfo{author}{Ament, L. J.~P.}, \bibinfo{author}{van Veenendaal, M.},
  \bibinfo{author}{Devereaux, T.~P.}, \bibinfo{author}{Hill, J.~P.} \&
  \bibinfo{author}{van~den Brink, J.}
\newblock \bibinfo{title}{{Resonant inelastic x-ray scattering studies of
  elementary excitations}}.
\newblock \emph{\bibinfo{journal}{Rev. Mod. Phys.}}
  \textbf{\bibinfo{volume}{83}}, \bibinfo{pages}{705--767}
  (\bibinfo{year}{2011}).

\bibitem{Jia2012}
\bibinfo{author}{Jia, C.~J.}, \bibinfo{author}{Chen, C.-C.},
  \bibinfo{author}{Sorini, A.~P.}, \bibinfo{author}{Moritz, B.} \&
  \bibinfo{author}{Devereaux, T.~P.}
\newblock \bibinfo{title}{{Uncovering selective excitations using the resonant
  profile of indirect inelastic x-ray scattering in correlated materials:
  observing two-magnon scattering and relation to the dynamical structure
  factor}}.
\newblock \emph{\bibinfo{journal}{New Journal of Physics}}
  \textbf{\bibinfo{volume}{14}}, \bibinfo{pages}{113038}
  (\bibinfo{year}{2012}).

\bibitem{WeiSheng}
\bibinfo{author}{Lee, W.~S.} \emph{et~al.}
\newblock \bibinfo{title}{{Asymmetry of collective excitations in electron and
  hole doped cuprate superconductors}}.
\newblock \emph{\bibinfo{journal}{http://arxiv.org/abs/1308.4740}}
  (\bibinfo{year}{2013}).

\bibitem{Vladimirov2009}
\bibinfo{author}{Vladimirov, A.~A.}, \bibinfo{author}{Ihle, D.} \&
  \bibinfo{author}{Plakida, N.~M.}
\newblock \bibinfo{title}{{Dynamic spin susceptibility in the $t\text{-}J$
  model}}.
\newblock \emph{\bibinfo{journal}{Phys. Rev. B}} \textbf{\bibinfo{volume}{80}},
  \bibinfo{pages}{104425} (\bibinfo{year}{2009}).

\bibitem{Kar2011}
\bibinfo{author}{Kar, S.} \& \bibinfo{author}{Manousakis, E.}
\newblock \bibinfo{title}{{Hole spectral functions in lightly doped quantum
  antiferromagnets}}.
\newblock \emph{\bibinfo{journal}{Phys. Rev. B}} \textbf{\bibinfo{volume}{84}},
  \bibinfo{pages}{205107} (\bibinfo{year}{2011}).

\bibitem{Daghofer2008}
\bibinfo{author}{Daghofer, M.}, \bibinfo{author}{Wohlfeld, K.},
  \bibinfo{author}{{Ole\ifmmode \acute{s}\else {\'s}\fi{}}, A.~M.},
  \bibinfo{author}{Arrigoni, E.} \& \bibinfo{author}{Horsch, P.}
\newblock \bibinfo{title}{{Absence of Hole Confinement in Transition-Metal
  Oxides with Orbital Degeneracy}}.
\newblock \emph{\bibinfo{journal}{Phys. Rev. Lett.}}
  \textbf{\bibinfo{volume}{100}}, \bibinfo{pages}{066403}
  (\bibinfo{year}{2008}).

\bibitem{Bala1995}
\bibinfo{author}{Ba\l{}a, J.}, \bibinfo{author}{{Ole\ifmmode \acute{s}\else
  \'{s}\fi{}}, A.~M.} \& \bibinfo{author}{Zaanen, J.}
\newblock \bibinfo{title}{{Spin polarons in the $t$-$t'$-$J$ model}}.
\newblock \emph{\bibinfo{journal}{Phys. Rev. B}} \textbf{\bibinfo{volume}{52}},
  \bibinfo{pages}{4597--4606} (\bibinfo{year}{1995}).

\bibitem{Dellanoy2009}
\bibinfo{author}{Delannoy, J.-Y.~P.}, \bibinfo{author}{Gingras, M. J.~P.},
  \bibinfo{author}{Holdsworth, P. C.~W.} \& \bibinfo{author}{Tremblay,
  A.-M.~S.}
\newblock \bibinfo{title}{{Low-energy theory of the
  $t$-${t}^{'}$-${t}^{''}$-$U$ Hubbard model at half-filling: Interaction
  strengths in cuprate superconductors and an effective spin-only description
  of ${\text{La}}_{2}{\text{CuO}}_{4}$}}.
\newblock \emph{\bibinfo{journal}{Phys. Rev. B}} \textbf{\bibinfo{volume}{79}},
  \bibinfo{pages}{235130} (\bibinfo{year}{2009}).

\bibitem{JPSJNeutronReview}
\bibinfo{author}{Fujita, M.} \emph{et~al.}
\newblock \bibinfo{title}{Progress in neutron scattering studies of spin
  excitations in high-${T}_{\mathrm{c}}$ cuprates}.
\newblock \emph{\bibinfo{journal}{Journal of the Physical Society of Japan}}
  \textbf{\bibinfo{volume}{81}}, \bibinfo{pages}{011007}
  (\bibinfo{year}{2012}).

\bibitem{Muschler2010}
\bibinfo{author}{Muschler, B.} \emph{et~al.}
\newblock \bibinfo{title}{{Electron interactions and charge ordering in CuO$_2$
  compounds}}.
\newblock \emph{\bibinfo{journal}{The European Physical Journal : Special
  Topics}} \textbf{\bibinfo{volume}{188}}, \bibinfo{pages}{131--152}
  (\bibinfo{year}{2010}).

\bibitem{Onose2004}
\bibinfo{author}{Onose, Y.}, \bibinfo{author}{Taguchi, Y.},
  \bibinfo{author}{Ishizaka, K.} \& \bibinfo{author}{Tokura, Y.}
\newblock \bibinfo{title}{{Charge dynamics in underdoped
  ${\mathrm{Nd}}_{2-x}{\mathrm{Ce}}_{x}{\mathrm{CuO}}_{4}$: Pseudogap and
  related phenomena}}.
\newblock \emph{\bibinfo{journal}{Phys. Rev. B}} \textbf{\bibinfo{volume}{69}},
  \bibinfo{pages}{024504} (\bibinfo{year}{2004}).

\bibitem{Scalapino2012}
\bibinfo{author}{Scalapino, D.~J.}
\newblock \bibinfo{title}{{A common thread: The pairing interaction for
  unconventional superconductors}}.
\newblock \emph{\bibinfo{journal}{Rev. Mod. Phys.}}
  \textbf{\bibinfo{volume}{84}}, \bibinfo{pages}{1383--1417}
  (\bibinfo{year}{2012}).

\bibitem{Tallon}
\bibinfo{author}{Mallett, B.} \emph{et~al.}
\newblock \bibinfo{title}{{${T}_{\mathrm{c}}$ is insensitive to magnetic
  interactions in high-${T}_{\mathrm{c}}$ superconductors}}.
\newblock \emph{\bibinfo{journal}{http://arxiv.org/abs/1202.5078}}
  (\bibinfo{year}{2012}).

\bibitem{White_Scalapino_1999}
\bibinfo{author}{White, S.~R.} \& \bibinfo{author}{Scalapino, D.~J.}
\newblock \bibinfo{title}{{Competition between stripes and pairing in a
  $t$-$t'$-$J$ model}}.
\newblock \emph{\bibinfo{journal}{Phys. Rev. B}} \textbf{\bibinfo{volume}{60}},
  \bibinfo{pages}{R753--R756} (\bibinfo{year}{1999}).

\bibitem{RevModPhys.79.175}
\bibinfo{author}{Devereaux, T.~P.} \& \bibinfo{author}{Hackl, R.}
\newblock \bibinfo{title}{{Inelastic light scattering from correlated
  electrons}}.
\newblock \emph{\bibinfo{journal}{Rev. Mod. Phys.}}
  \textbf{\bibinfo{volume}{79}}, \bibinfo{pages}{175--233}
  (\bibinfo{year}{2007}).

\bibitem{Blank1981}
\bibinfo{author}{Blankenbecler, R.}, \bibinfo{author}{Scalapino, D.~J.} \&
  \bibinfo{author}{Sugar, R.~L.}
\newblock \bibinfo{title}{{Monte Carlo calculations of coupled boson-fermion
  systems. I}}.
\newblock \emph{\bibinfo{journal}{Phys. Rev. D}} \textbf{\bibinfo{volume}{24}},
  \bibinfo{pages}{2278--2286} (\bibinfo{year}{1981}).

\bibitem{White1989}
\bibinfo{author}{White, S.~R.} \emph{et~al.}
\newblock \bibinfo{title}{{Numerical study of the two-dimensional Hubbard
  model}}.
\newblock \emph{\bibinfo{journal}{Phys. Rev. B}} \textbf{\bibinfo{volume}{40}},
  \bibinfo{pages}{506--516} (\bibinfo{year}{1989}).

\bibitem{Jarrell1996}
\bibinfo{author}{Jarrell, M.} \& \bibinfo{author}{Gubernatis, J.}
\newblock \bibinfo{title}{{Bayesian inference and the analytic continuation of
  imaginary-time quantum Monte Carlo data}}.
\newblock \emph{\bibinfo{journal}{Physics Reports}}
  \textbf{\bibinfo{volume}{269}}, \bibinfo{pages}{133--195}
  (\bibinfo{year}{1996}).

\bibitem{MEM2}
\bibinfo{author}{Macridin, A.}, \bibinfo{author}{Doluweera, S.~P.},
  \bibinfo{author}{Jarrell, M.} \& \bibinfo{author}{Maier, T.}
\newblock \bibinfo{title}{{Analytic continuation of QMC data with a sign
  problem}}.
\newblock \emph{\bibinfo{journal}{http://arxiv.org/abs/cond-mat/0410098}}
  (\bibinfo{year}{2004}).

\bibitem{SignProblemPRB}
\bibinfo{author}{Loh, E.~Y.} \emph{et~al.}
\newblock \bibinfo{title}{Sign problem in the numerical simulation of
  many-electron systems}.
\newblock \emph{\bibinfo{journal}{Phys. Rev. B}} \textbf{\bibinfo{volume}{41}},
  \bibinfo{pages}{9301--9307} (\bibinfo{year}{1990}).

\bibitem{Tsutsui2000}
\bibinfo{author}{Tsutsui, K.}, \bibinfo{author}{Kondo, H.},
  \bibinfo{author}{Tohyama, T.} \& \bibinfo{author}{Maekawa, S.}
\newblock \bibinfo{title}{{Resonant inelastic X-ray scattering spectrum in
  high-${T}_{\mathrm{c}}$ cuprates}}.
\newblock \emph{\bibinfo{journal}{Physica B: Condensed Matter}}
  \textbf{\bibinfo{volume}{284--288, Part 1}}, \bibinfo{pages}{457--458}
  (\bibinfo{year}{2000}).

\bibitem{Kourtis2012}
\bibinfo{author}{Kourtis, S.}, \bibinfo{author}{van~den Brink, J.} \&
  \bibinfo{author}{Daghofer, M.}
\newblock \bibinfo{title}{{Exact diagonalization results for resonant inelastic
  x-ray scattering spectra of one-dimensional Mott insulators}}.
\newblock \emph{\bibinfo{journal}{Phys. Rev. B}} \textbf{\bibinfo{volume}{85}},
  \bibinfo{pages}{064423} (\bibinfo{year}{2012}).

\bibitem{Harris1967}
\bibinfo{author}{Harris, A.~B.} \& \bibinfo{author}{Lange, R.~V.}
\newblock \bibinfo{title}{{Single-Particle Excitations in Narrow Energy
  Bands}}.
\newblock \emph{\bibinfo{journal}{Phys. Rev.}} \textbf{\bibinfo{volume}{157}},
  \bibinfo{pages}{295--314} (\bibinfo{year}{1967}).

\bibitem{Millis2012}
\bibinfo{author}{Lin, N.}, \bibinfo{author}{Gull, E.} \&
  \bibinfo{author}{Millis, A.~J.}
\newblock \bibinfo{title}{{Two-Particle Response in Cluster Dynamical
  Mean-Field Theory: Formalism and Application to the Raman Response of
  High-Temperature Superconductors}}.
\newblock \emph{\bibinfo{journal}{Phys. Rev. Lett.}}
  \textbf{\bibinfo{volume}{109}}, \bibinfo{pages}{106401}
  (\bibinfo{year}{2012}).

\bibitem{Bisogni2012}
\bibinfo{author}{Bisogni, V.} \emph{et~al.}
\newblock \bibinfo{title}{{Bimagnon studies in cuprates with resonant inelastic
  x-ray scattering at the $O$ $K$-edge. $II$. Doping effect in
  La$_{2-x}$Sr$_{x}$CuO$_4$}}.
\newblock \emph{\bibinfo{journal}{Phys. Rev. B}} \textbf{\bibinfo{volume}{85}},
  \bibinfo{pages}{214528} (\bibinfo{year}{2012}).

\bibitem{Bisogni2013}
\bibinfo{author}{{Bisogni}, V.} \emph{et~al.}
\newblock \bibinfo{title}{{Femtosecond dynamics of magnetic excitations from
  resonant inelastic x-ray scattering in CaCu$_2$O$_3$}}.
\newblock \emph{\bibinfo{journal}{http://arxiv.org/abs/1307.8393}}
  (\bibinfo{year}{2013}).

\bibitem{Igarashi2012}
\bibinfo{author}{Igarashi, J.-i.} \& \bibinfo{author}{Nagao, T.}
\newblock \bibinfo{title}{{Magnetic excitations in $L$-edge resonant inelastic
  x-ray scattering from cuprate compounds}}.
\newblock \emph{\bibinfo{journal}{Phys. Rev. B}} \textbf{\bibinfo{volume}{85}},
  \bibinfo{pages}{064421} (\bibinfo{year}{2012}).
\newblock \urlprefix\url{http://link.aps.org/doi/10.1103/PhysRevB.85.064421}.

\bibitem{Martinez1991}
\bibinfo{author}{Martinez, G.} \& \bibinfo{author}{Horsch, P.}
\newblock \bibinfo{title}{{Spin polarons in the t-J model}}.
\newblock \emph{\bibinfo{journal}{Phys.~Rev.~B}} \textbf{\bibinfo{volume}{44}},
  \bibinfo{pages}{317--331} (\bibinfo{year}{1991}).

\end{thebibliography}


\end{document}